\documentstyle{article}
\def\complex{{\kern .1em {\raise .47ex \hbox
{$\scriptscriptstyle
|$}}
\kern -.4em {\rm C}}}
\def\be{\begin{equation}} \def\ee{\end{equation}} 
\def\ba{\begin{array}} \def\ea{\end{array}} \def\ds{\displaystyle} \def%
\tr{\mbox{Tr}} 
\title{Topics in Quantum Dynamics} 
\author{A.  Jadczyk \\
Institute of Theoretical Physics, 
 University of Wroc{\l }aw,  \\
Pl.  Maxa Borna 9,  PL 50 204 Wroc{\l }aw,  Poland} 
\begin{document}
\newtheorem{theorem}{Theorem}
\maketitle
\section{Introduction}
\subsection{The Two Kinds of Evolution}
In these lectures I will discuss the two kinds of evolution of
quantum systems:  \begin{itemize}
\item CONTINUOUS evolution of closed systems
\item STOCHASTIC evolution of open systems.  
\end{itemize}
The first type
concerns evolution of {\sl closed, 
isolated}\footnote{
{\sl Emphasized}\,  style will be  used in these notes
for  concepts that are {\sl
important},  but will not be {\sl explained}.  Sometimes explanation
would need too much space,  but sometimes because these are either {\sl
primitive} or {\sl meta--language} notions. } quantum systems that evolve
under the action of prescribed external forces,  but are not disturbed by
observations,  and are not coupled thermodynamically or in some other
irreversible way to the environment.  This evolution is governed by the
Schr\"odinger equation and is also known as a {\sl unitary} or,  more
generally,  as an {\sl automorphic} evolution.  In contrast to this idealized
case  (only approximately valid, when irreversible effects can be neglected),
quantum theory is also concerned with a different kind of change of state. 
It was first formulated by J.  von Neumann  (cf.  \cite[Ch.  V. 1]{neu} and is
known as von Neumann -- L\"uders {\sl projection postulate}.  It tells us
roughly this:  if some quantum mechanical observable is being measured,
then - as
a consequence of this measurement - the actual state of the quantum system 
{\sl jumps} into one of the eigenstates of the measured
observable. \footnote{L\"uders
\cite{lud} noticed that this formulation is ambiguous in case of
degenerate eigenvalues,  and generalized it to cover also this
situation. } This jump was thought to be abrupt and take no time at all
,  it is also known as {\sl
reduction of the wave packet}.  Some physicists feel quite uneasy about this
von Neumann's postulate,  to the extent that they reject it either as too
primitive  (not described by dynamical equations) or as unnecessary.  We will
come to this point later,  in Chapter 3,  when we will discuss {\sl piecewise 
deterministic}
stochastic processes that unite both kinds of evolution. 

\subsection{The Schr\"odinger Equation}

It is rather easy to explain to the mathematician the {\sl Dirac} equation --
it became already a part of the mathematical folklore.  But Dirac's equation
belongs to {\sl field theory} rather than to
{\sl quantum theory}. \footnote{In
these lectures,  "quantum theory" usually means
"quantum mechanics",  although much of the concepts that we discuss
are applicable also to systems with infinite number of degrees of
freedom and in particular to quantum field theory. } Physicists are being
taught in the course of their education rather early that every attempt at a
sharp localization of a relativistic particle results in creation and
annihilation processes.  Sometimes it is phrased as:  "there is no relativistic
quantum mechanics -- one needs to go directly to Relativistic Quantum Field
Theory".  Unfortunately we know of none non-trivial,  finite,  
relativistic quantum
field theory in four dimensional space--time continuum.  Thus we are left
with the mathematics of {\sl perturbation} theory.  Some physicists believe
that the physical ideas of relativistic quantum field theory are sound,  that
it is the best theory we ever had,  that it is "the exact description of 
nature",  that
the difficulties we have with it are only temporary,  and that they will be 
overcomed one day --
the day when bright mathematicians will provide us with new,  better,  more
powerful tools.  Some other say:  perturbation theory is more than sufficient
for all practical purposes,  no new tools are needed,  that is how {\sl physics
is} -- so mathematicians better accept it,  digest it,  and help the physicists
to make it more rigorous and to understand what it is really about.  Still some
other\footnote{Including the  author of these notes. },  a
minority,  also believe that it is only a {\sl temporary} situation,  which
one day will be resolved.  But the resolution will come
owing to essentially
\underline{new physical ideas}, 
and it will result in a new quantum paradigm,  more appealing than the
present one.  It should perhaps be not a surprise if,  in an appropriate sense,  
all
these points of view will turn out to be right.  In these lectures we will be
concerned with the well established Schr\"odinger equation,  which is at the
very basis of the current quantum scheme,  and with its dissipative
generalization -- the Liouville equation. 
In these equations we assume that we 
know what the {\sl time} is.  Such a knowledge is negated in special
relativity, \footnote{There exists however so called "relativistic
Fock--Schwinger proper time formalism" \cite[Ch. 2--5--4]{itz} where
one writes Schr\"odinger equation with Hamiltonian replaced by
"super--Hamiltonian,  and time replaced by "proper time"} and this results
in turn in all kinds of troubles that we are facing since the birth of
Einstein's relativity till this day.
\footnote{One could try to "explain" time by saying that there is a
preferred time direction selected by
the choice of a thermal state of the universe. But that is of no help
at all, until
we are told how it happens that a particular thermal state is
being achieved.}
\\ The Schr\"odinger equation is more
difficult than the Dirac one,  and this for two reasons: 
first,  it lives on the
background of {\sl Galilean} relativity -- which have to deal with much more
intricate geometric structures than Einstein relativity. \footnote{
In group theoretical terms:  the proper Lorentz group is simple, 
while the proper Galilei group is not. } 
 Second, 
Schr\"odinger's equation is about {\sl Quantum Mechanics} and we have to take
care about {\sl probabilistic interpretation},  {\sl observables},  {\sl states
} etc.  -- which is possible for Schr\"odinger equation but faces
problems in the first--quantized Dirac theory.

Let me first make a general comment about Quantum Theory.  There are
physicists who would say:  quantum theory is about calculating of {\sl 
Green's Functions} -- all numbers of interest can be obtained from these
functions and all the other mathematical constructs usually connected with 
quantum theory are
superfluous and unnecessary! It is not my intention to depreciate the 
achievements of Green's
Function Calculators.  But for me quantum theory -- 
as any other physical theory -- should be
about  {\sl explaining} things that happen outside of us -- as long as such {\sl 
explanations} are possible.  The situation in Quantum Theory today,  more than
60 years after its birth,  is that Quantum Theory {\sl explains much less}
than we would like to have been explained.
To reduce quantum theory to Green's
function calculations  is to reduce its explanatory power almost
to  zero.  
 It may of course be that in the coming 
  Twenty First  Century humanity will unanimously
recognize the fact that `understanding'\thinspace was a luxury,  a luxury of
the "primitive age" that is gone for ever.  But I think it is worth
while to take a chance and to try to  understand as much as can be
understood in a given situation.  Today we are trying to understand Nature
in terms of {\sl geometrical pictures} and
{\sl random processes}\footnote{The paradigm  may however change in no so
distant  future -- we may soon try to understand the Universe as a
{\sl computing machine},  with geometry replaced by geometry of
connections,  and randomness replaced by a variant of algorithmic
complexity. } More specifically,  in quantum theory,  we are trying to
understand in terms of {\sl observables},  {\sl states}, 
{\sl complex Feynman amplitudes}
etc.  In the next chapter,  we will show the way that leads to the
 Schr\"odinger Equation using
{\sl geometrical language} as much as possible.  However,  we will not use the
machinery of {\sl geometrical quantization} because it treats {\sl time}
simply as a {\sl parameter},  and {\sl space} as absolute and
given once for all. \footnote{Cf.  e. g.  Ref.  \cite[Ch. 9]{snia}. }
On the other hand,  geometrical quantization introduces many
advanced tools
that are unnecessary
for our purposes,  while at the same time it lacks the concepts which are
important and necessary. \\ 
Before entering the subject let me tell you the
distinguishing feature of the approach that I am advocating,  and that
will be sketched below in Ch. 
2:  one obtains a fibration of Hilbert spaces ${\cal H}=\bigcup {\cal H}
_t$ over time.  There is  a distinguished family of local
trivializations,  a family parameterized by 
\begin{itemize}
\item  space--time observer 
\item $U (1)$ gauge.
\end{itemize}
For each $t$, the Hilbert space ${\cal H}_t$ is a Hilbert space
of sections of a complex line bundle over $E_t$. A space--time observer
(that is, a reference frame)
allows us to identify the spaces $E_t$ for different $t$-s, while
a $U(1)$ gauge allows us to identify the fibers.
Schr\"odinger's dynamics of a particle in external gravitational and
electromagnetic
fields
is given by a {\sl Hermitian connection} in ${\cal H}. $ Solutions of
Schr\"odinger
equation are parallel sections of ${\cal H}. $ Thus Schr\"odinger equation
can be written
as \be
\nabla \Psi = 0 \ee
or,  in a local trivialization,  as \be
{\frac{{\partial \Psi  (t)}}{{\partial t}}} + 
{i\over\hbar}H (t)\Psi  (t) =
0,  \ee
where $H (t)$ will be a self--adjoint operator in ${\cal H}_t$. 
\footnote{A similar idea was mentioned in \cite{aso}.  For a detailed
description of all the constructions -- see the forthcoming book \cite{jamo}}
Gravitational and electromagnetic forces are
coded into this Schr\"odinger's connection. 
Let us discuss the resulting structure.  First of all
there is not a single  Hilbert space but a
\underline{family} of Hilbert spaces. 
These Hilbert spaces can be {\sl identified} either using an
{\sl observer}  and
{\sl gauge}  or,better, by using a background dynamical connection.
It is only after so doing 
that one arrives at
single--Hilbert--space picture of the textbook Quantum Mechanics --
a picture that
is the constant source of lot of confusion. \footnote{Notable
exceptions can be found in  publications from the
Genev\'e school of Jauch and Piron. }\\ 
In Quantum Mechanics we have a {\sl dual}
scheme -- we use the concepts of {\sl observables} and {\sl states}.  We often
use the word {\sl measurement} in a mutilated sense of simply {\sl pairing}
an observable $a$ with a state $\phi $ to get the {\sl expected result} -- 
a {\sl number} $
<\phi , a>. $ \\ One comment is in place here:  to compare the results of actual
measurements with predictions of the theory -- one needs only {\sl real}
numbers.  However experience proved that quantum theory with
only--real--numbers is inadequate.  So,  even if the fundamental role of
$\sqrt{-1}$ in Quantum Theory is far from being fully understood --
we use in Quantum Theory only {\sl complex} Hilbert spaces,  complex algebras
etc.\footnote{Quaternionic  structures,  on the other hand,  can be always 
understood as complex
ones with an extra structure -- they are unnecessary.  } 
However,  usually,  only real numbers are at the end interpreted.  
 
 Now,  it is
not always clear what is understood by {\sl states} and {\sl observables}. 
There are several possibilities:  
\begin{figure}[hbt]
\unitlength=0.8mm
\begin{picture}(100, 120)
\put(0,90){\bf STATE}
\put(70,90){\bf OBSERVABLE}
\put(0,60){\framebox(30,20){INSTANT}}
\put(70,60){\framebox(30,20){INSTANT}}
\put(0,10){\framebox(30,20){PARALLEL}}
\put(70,10){\framebox(30,20){{\shortstack{PARALLEL\\ or \\ARBITRARY}}}}
\put(30,70){\line(4,-5){40}}
\put(70,70){\line(-4,-5){40}}
\put(32,70){--state}
\put(32,20){--section}
\put(102,20){--section}
\end{picture}
\caption{There are several possibilities of understanding 
of state and observables. They can be {\sl instant}\ , and thus
time--dependent, or they can be {\sl time--sections} -- thus time--
independent}
\end{figure}
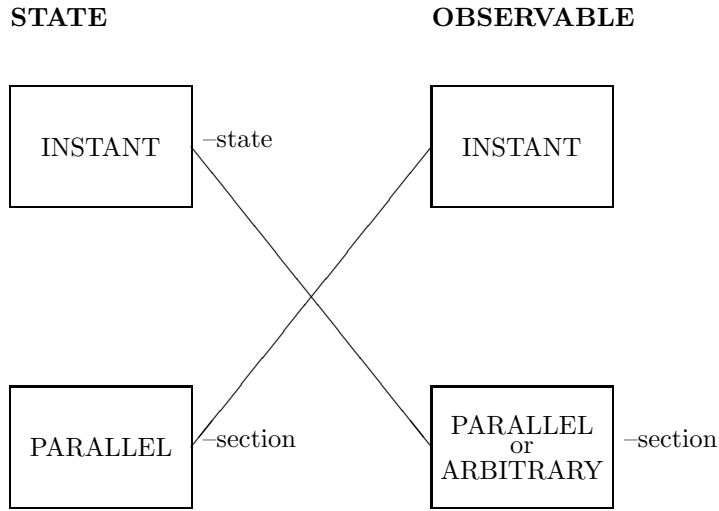

As it was already said,  it is usual in standard presentations of the quantum
theory to {\sl identify}\ the Hilbert spaces ${\cal H}_t$.  There are
several
options there.  Either we identify them according to an observer  (+ gauge) or
according to the dynamics.  If we identify according to the actual dynamics,
then
states do not change in time -- it is always the same state--vector,  but
observables  (like position coordinates) do change in time -- we have what is
called the Heisenberg picture.  If we identify them according to some
background "free" dynamics -- we have so called {\sl interaction
picture.} Or,  we can identify Hilbert spaces according
to an observer -- then observables do not change in time,  but state vector
is changing -- we get the Schr\"odinger picture.  
$$
\begin{array}{c}
\cr
\hbox{\LARGE IDENTIFICATION}
\cr\cr
{\left\{ 
\begin{array}{lll}
\hbox{according to the dynamics}
\qquad & \Longrightarrow
\qquad &\hbox{Heisenberg picture} \cr\cr
\hbox{according to an observer}
& \Longrightarrow
& \hbox{Schr\"{o}dinger picture}
\end{array} \right. 
}
\cr\cr
\end{array}
$$
However,  there is no reason at all to identify the ${\cal H}_t$-s.  Then
dynamics is given by parallel transport operators:  
$$
U_{t, s}: {\sl H}_s\rightarrow {\sl H}_t, \hspace{1cm}s\leq t 
$$
$$
U_{t, s}U_{s, r}=U_{t, r} 
$$
$$
U_{t, t}=id_{{\cal H}_t}.  
$$

\subsection{Dissipative Dynamics}

The Schr\"odinger equation describes time evolution of pure states of a quantum
system,  for instance evolution of pure states of a quantum particle, 
 or of a many body system.  
Even if these states
contain only {\sl statistical information} about most of the physical
 quantities,  the
Schr\"odinger evolution of pure states is {\sl continuous and deterministic}
.  Under this evolution Hilbert space vector representing the actual state of
the system\footnote{Some physicists deny "objectivity" of quantum states
-- they would say that Hilbert space vectors describe not
states of the system,  but states of {\sl knowledge} or {\sl
information} about  the system.  In a recent series of
papers  (see \cite{aha} and references therein)
Aharonov and Vaidman \cite{aha}
attempt to justify
objectivity of quantum states.  Unfortunately their arguments contain
a loophole. } changes
continuously with time,  and with it there is a continuous evolution of {\sl 
probabilities} or {\sl potentialities},  but nothing {\sl happens} -- the
formalism leaves no place for {\sl events}.  Schr\"odinger equation helps us
very little,  or nothing at all,  to understand {\sl how} potential becomes
real.  So,  if we aim at understanding of this {\sl process of becoming},  if we
want to describe it by mathematical equations and to simulate it with
computers -- we must go beyond Schr\"odinger's dynamics.  As it happens,  we
do not have to go very far -- it is sufficient to relax only one  (but {\sl 
important}) property of Schr\"odinger's dynamics and to admit that pure
states can evolve into mixtures.  Instead of Schr\"odinger equation we have
then a so called Liouville equation that describes time evolution of mixed
states.  It contains Schr\"odinger
equation as a special case.\footnote{
It should be noted,  however,  that Schr\"odinger equation
describes evolution of state {\sl vectors}.  and thus contains direct
information about {\sl phases}.  This information is absent in
the Liouville equation,  and its restoration  (e. g.  as it is with the
{\sl Berry phase}) 
may sometimes create  a non--trivial task. }
It was shown in \cite{bla5} that using the Liouville type of
dynamics it is possible to describe coupling between quantum
systems and classical degrees of freedom of measurement devices. 
One can derive also a piecewise deterministic random process
that takes place on the manifold of pure states.  In  this way
one obtains a minimal description of "quantum jumps"  (or
"reduction of wave packets") and accompanying,  directly
observable jumps of the coupled classical devices.  In Ch.  3
simple models of such couplings will be discussed.  The
interested reader will find more examples in Refs. 
\cite{bla2}--\cite{bla5}.\footnote{Cf. also the recent (June 1994) paper "Particle Tracks,
Events and Quantum Theory", by the author.}
In particular in \cite{bla5} the most
advanced model of this kind,  the SQUID--tank
model is discussed in details.
\section{Geometry of Schr\"odinger's Equation}
\subsection{Preliminaries}
Galilean General Relativity is a theory of space--time
structure,  gravitation and electromagnetism based on the
assumption of existence of an absolute time function.  Many of the
cosmological models based on Einstein`s relativity admit also a 
distinguished time function.  Therefore Galilean physics is not
evidently wrong.  Its predictions must be tested by experiments. 
Galilean relativity is not
that elegant as the one of Einstein.  This can be already seen
from the group structures:  the homogeneous Lorentz group is
{\sl simple},  while the homogeneous Galilei group is a semidirect
product of the rotation group and of three commuting boosts.
Einstein`s theory of gravitation is based on one metric tensor, 
while Galilean gravity needs both:  space metric and space--time
connection.  Similarly for quantum mechanics:  it is rather
straightforward to construct generally covariant wave equations
for Einstein`s relativity,  while general covariance and
geometrical meaning of the Schr\"{o}dinger equation was causing
problems,  and it was not discussed in textbooks.  In the
following sections we will present a brief overview of some
of these problems. 
\subsection{Galilean General Relativity}
Let us discuss briefly geometrical data that are needed for building up
generally covariant Schr\"odinger's equation.  More details can be
found in Ref. \cite{jamo}.\footnote{The reader may also consult \cite{duv}, 
where a different approach,  using dimensional reduction along a
null Killing vector,  is discussed.  }\\
Our space--time will be a refined version of that of Galilei and
of Newton,  i.e.  space--time with absolute simultaneouity. 
Four dimensional space--time $E$ is fibrated over one--dimensional
time $B . $ The fibers $E_t$ of $E$ are three--dimensional
Riemannian manifolds,  while the basis $B$ is an affine space over
${\bf R} . $ By a coordinate system on $E$ we will always mean a
coordinate system $x^\mu= (x^0, x^i) , $ $i=1, 2, 3\,  , $ {\sl adapted
to the
fibration.}  That means: any two events with the same coordinate $x^0$ are
simultaneous,  i.e.  in the same fibre of $E . $\\
Coordinate transformations between any two adapted coordinate systems
are of the form: 
$$
x^{0'}=x^0+const , 
$$
$$
x^{i'}=x^{i'}\left (x^0, x^{i}\right) . 
$$
We will denote by $\beta$ the time form $dx^0 . $ Thus in adapted
coordinates $\beta_0=1, \beta_i=0$. \\
$E$ is equipped with a contravariant {\sl degenerate}\, metric tensor which,
in adapted coordinates, takes the form
\def\linie{\vrule height 15pt depth 5pt}
\def\back{\noalign{\vskip-3pt}}
$$\pmatrix{0 &\linie & \phantom{0 }& 0 &\phantom{0}\cr
\noalign{\hrule}
&\linie && \cr\back
0 &\linie &  & g^{ij} & 
\cr\back
& \linie && } \ . $$
where $g^{ij}$ is of signature $ (+++) . $ We denote by $g_{ij}$ the
inverse $3\times 3$ matrix.  It defines Riemannian metric in the
fibers of $E . $\\
We assume a torsion--free connection in $E$ that preserves
the two geometrical objects $g^{\mu\nu}$ and $\beta$.
\footnote{Some of these assumptions are superfluous
as they would
follow anyhow from the assumption $d\Omega=0$ in the next
paragraph. }
The condition $\nabla\beta =0$ is equivalent to the conditions 
$\Gamma^0_{\mu\nu}=0$ on the connection coefficients.  Let us
introduce the notation $\Gamma_{\mu\nu, i}=g_{ij}\Gamma^j_{\mu\nu}. $
Then $\nabla g^{\mu\nu}=0$ is equivalent to the equations: 
\be
\partial_\mu g_{ij}=\Gamma_{\mu i, j}+\Gamma_{\mu j, i} . 
\ee
Then,  because of the assumed zero torsion,  the space part of the
connection can be expressed in terms of the space metric in
the Levi-Civita form:
\be
\Gamma_{ij, k}={1\over2}\left ( \partial_i g_{jk}+
\partial_j g_{ik}-
\partial_k g_{ij}\right) . 
\ee
>From the remaining equations:
\be
\partial_0 g_{ij}= \Gamma_{0i, j} + \Gamma_{0j, i} 
\ee
we find that the  $(i,j)$--symmetric part of $\Gamma_{0i, j}$ is equal to
${1\over2}\partial_0 g_{ij} , $ otherwise the connection is
undetermined.  We can write it, introducing a new
geometrical object $\Phi$, as
\be
\Gamma_{i0, j}={1\over2}\left (\partial_0 g_{ij}+\Phi_{ij}\right) , 
\ee
\be
\Gamma_{00, j}=\Phi_{0j} , 
\ee
where $\Phi_{\mu\nu}=-\Phi_{\mu\nu}$ is antisymmetric. Notice
that $\Phi$ is not a tensor, except for pure space transformations
or time translations.
\subsubsection{The Bundle of Galilei Frames}
A basis $e_{\mu}$ in $TE$ is called a Galilei frame if
$e_0=\partial_0$,  and if $e_i$ are unit space--like vectors.  If
$e_\mu$ and $\tilde{e}_\mu$ are two Galilei frames at the same
space--time point,  then they are related by a transformation of the homogeneous
Galilei group $G$: 
\be
\tilde{e}_0=e_0+{\bf e\cdot v} , 
\ee
\be
{\bf \tilde{e}}={\bf e}{\bf\Lambda} , 
\ee
where ${\bf v}\in {\bf R}^3$ and ${\bf \Lambda}$ is an orthogonal
$3\times 3$ matrix. 
The bundle of Galilei frames is a principal $G$ bundle. 
\subsubsection{The Bundle of Velocities }
The homogeneous Galilei group $G$ acts on ${\bf R}^3$ in two
natural ways:  by linear and by affine transformations.  The first
action is not {\sl effective} one -- it involves only the rotations:
\be
 ({\bf \Lambda}, {\bf v}): {\bf x}\mapsto {\bf\Lambda x}. 
\ee
The bundle associated to this action can be identified with the
vertical subbundle of $TE$ -- i.e.  with the bundle $VE$ of vectors
tangent to the fibers of $E\rightarrow B$. \\
$G$ acts also on ${\bf R}^3$ by affine
isometries : 
\be
 ({\bf \Lambda}, {\bf v}): {\bf y}\mapsto {\bf\Lambda y}+{\bf v}  . 
\ee
To this action there corresponds an associated bundle,  which is an
affine bundle over the vector bundle $VE$. 
 It can be identified with the subbundle of $TE$ consisting of
vectors $\xi$ tangent to $E$, and such that $\beta  (\xi )=1$ or,
equivalently,  as the bundle of {\sl first jets of sections}\, of
$E\rightarrow B$.  We will call it  $J_1E$. \\
We will denote by $ (x^0, x^{i}, y_0^{i})$ the coordinates in $J_1E$
corresponding to coordinates $x^{\mu}$  of $E$.  
\subsection{The Presymplectic Form }
The connection $\Gamma$ can be also considered as a principal
connection in the bundle of Galilei frames.  It induces an affine
connection in the affine bundle $J_1E\stackrel{\pi}{\longrightarrow}E$.  
As
a result,  it defines a natural $VE$--valued one--form $\nu_\Gamma$ on
$J_1E$.  It can be described as follows:  given a vector $\xi$ tangent
to $J_1E$ at
$ (x, y_0)$ it projects onto $d\pi  (\xi )$.  Then $\nu_\Gamma  (\xi) $
is defined as the difference of $\xi$ and the horizontal lift of
$d\pi  (\xi )$.  It is a vertical tangent vector to $J_1E$,  and can be
identified with an element of $VE$.  In coordinates: 
\be
\nu_\Gamma^{i}=dy_0^i +\left (\Gamma_{\mu j}^iy_0^j+\Gamma_{\mu
0}^i \right)dx^\mu . 
\ee
There is another $VE$--valued one--form on $J_1E$,  namely the
canonical form $\theta$.  Given $\xi$ at $ (x, y_0)$,  we can decompose
$\xi$ into space-- and time--component along $y_0$.  Then $\theta
 (\xi)$ is defined as its space component.  In coordinates: 
\be
\theta^i=dx^i -y_0^i \, dx^0 . 
\ee
Then,  because the fibers of $VE$ are endowed with metric
$g^{mn}$,  we can build out the following important two--form
$\Omega$ on $J_1E$: 
\be
\Omega=g_{mn}\nu_\Gamma^m\wedge\theta^n . \ee
Explicitly
\be
\begin{array}{r}
\Omega = g_{lm} \left[ dy^{l}_{0} \wedge
\theta^{m} + 
\left ( \Gamma^{l}_{jk} 
y^{k}_{0}
+
\Gamma^{l}_{j0} \right)
\theta^{j}\wedge \theta^{m}
\  + \right. 
\cr\cr
\left. 
 + \  \left (  
 \Gamma^{l}_{jk} 
y^{j}_{0}y^{k}_{0}
+
\Gamma^{l}_{0k} 
y^{k}_{0}
+
\Gamma^{l}_{k0} 
y^{k}_{0}
+
\Gamma^{l}_{00}
\right)
dx^{0}\wedge\theta^{m}
\right]
\end{array}
\ee
The following theorem,  proven in \cite{jamo},  gives a necessary and
sufficient condition for
$\Omega$ to be closed. 
\begin{theorem}
The following conditions  (i--iii) are equivalent: \\
\begin{description}
\item[ (i)\ \ ] $d\Omega=0 , $ 
\item[ (ii)\ ] $R^{\mu \phantom{\nu}  
\sigma}_{\phantom{\mu}  \nu \phantom{\sigma}  \rho} =
R^{\sigma\phantom{\rho}  \mu}_{ \phantom{\sigma} \rho\phantom{\mu} \nu}$
where $R_{\mu\nu \phantom{\sigma}  \rho}^{\phantom{\mu \nu} \sigma}$ 
is the curvature tensor of
$\Gamma$ and $R^{\mu\ \phantom{\nu}\sigma}_{\phantom{\mu}
 \nu \phantom{\sigma} \rho}
 =g^{\mu\lambda}
R_{\lambda \nu \phantom{\sigma} \rho}^{\phantom{\mu \nu}\sigma} $
\item[ (iii)] $\partial_{[ \mu} \Phi_{\nu\sigma]}=0 $.
\footnote{Notice that because $\Phi$ is not a tensor, the last
condition need not be, a priori, generally covariant.}
\end{description}
\end{theorem}    
\subsection{Quantum Bundle and Quantum Connection}

Let $Q$ be a principal $U (1)$ bundle over $E$
and let $Q^{\uparrow}$ be its pullback to $J_{1}E$. 
We denote by $P$ and $P^{\uparrow}$ the associated Hermitian
line bundles corresponding to the natural action of $U (1)$ on 
$\complex$.  There is a special class of principal connections on
$Q^{\uparrow}$,  namely those whose connection forms vanish on
vectors tangent to the fibers of $Q^{\uparrow} \to
 Q$.  As has
been discussed in \cite{mod}
specifying such a connection on
$Q^{\uparrow}$ is equivalent to specifying a {\sl system of
connections} on $Q$ parameterized by the points in the fibers of
$Q^{\uparrow} \to Q$.  Following the terminology of \cite{mod} we
call such a connection {\sl universal}.\footnote{i.e.  {\sl
universal} for the system of connections.}
The fundamental assumption that leads to the Schr\"{o}dinger
equations  reads as follows: 
\\
{\bf Quantization Assumption: }
{\sl There exists a universal connection
$\omega$  $Q^{\uparrow}$ whose curvature is
$i\Omega$.}\footnote{We choose the physical units in such a way
that the Planck constant
$\hbar$ and mass of the quantum particle $m$ are equal to 1. }

We call such an $\omega$ a {\sl quantum connection}. 
>From the explicit
form of $\Omega$ one can easily deduce that $\omega$ is
necessarily of the form
$$
\omega = i\left ( d \phi + a_{\mu} dx^{\mu}\right)
$$
where $0\leq \phi \leq 2\pi$ parameterizes the fibres of $Q$, 
$$
\begin{array}{l}
a_{0} = - {1\over 2} \ y^{2}_{0} + \alpha_{0} ,  \cr\cr
a_{i} =  g_{ij} y^{j}_{0} + \alpha_{i} ,  
\end{array}
$$
and $\alpha^{\nu} = \left (\alpha^{0}, \alpha^{i}\right)$ is a local
potential for $\Phi$. 
\subsection{Schr\"{o}dinger's Equation and Schr\"{o}dinger's Bundle}

As it is shown in \cite{jamo},  there exists a natural $U (1)$--invariant
metric on $P$ of signature $ (++++-)$.  Explicitly 
$$
\begin{array}{ll}
\displaystyle\mathop{g}\limits^{5}
&  =\   d\phi \otimes dx^{0} + dx^{0} \otimes d\phi
\cr\cr
& +g_{ij}dx^{i}\otimes dx^{j} + 2 \alpha_{0}
dx^{0}\otimes  dx^{0} + \cr\cr
& + \alpha_{i}
\left ( dx^{i}\otimes dx^{0} + dx^{0}\otimes dx^{i} \right) . 
\end{array}
$$
Using this metric we can build out a natural Lagrangian for
equivariant functions $\psi :  P \rightarrow \complex$ or, 
equivalently,  for sections of the line bundle $Q$.  The Euler--Cartan
equation for this Lagrangian will prove to be  nothing but the
Schr\"{o}dinger equation.  Notice that the action of $U (1)$ group
on $P$ defines
an Killing vector field for $\mathop{g}\limits^{5}$ which is isotropic.
Therefore the above construction can explain why the
approach of \cite{duv} works.

More precisely,  the construction leading  to the generally
covariant Schr\"{o}\-din\-ger--Pauli equation for a charged spin $1/2$
particle in external gravitational and electromagnetic field can
be described as follows. 

The contravariant metric $\mathop{g}\limits^{5}{}^{-1}
=\left ( g^{\alpha \beta}\right)$, 
$\alpha,  \beta = 0, 1, 2, 3, 5$, 
\be
\left ( g^{\alpha \beta} \right)
= \pmatrix{ 0 & 0 & 1 \cr\cr
0 & g^{ij} & -g^{ij}a_{j} \cr\cr
1 & -g^{ij}a_{j} & \b{a}^{2}-2a_{0} }
\ . 
\ee
can be obtained from the following Clifford algebra of $4\times 4$
complex matrices: 
\be
\begin{array}{l}
\gamma^{0} = \pmatrix{0 & 0 \cr
1 & 0} 
\cr\cr
\gamma^{i} = \pmatrix{ \sigma^{i} & 0 \cr
0 & - \sigma^{i} }
\cr\cr
\gamma^{5} = \pmatrix{-\underline{\sigma}\cdot \underline{a} & 1 \cr\cr
 -2a_{0} & \underline{\sigma} \cdot \underline{a} }
\end{array}
\ee
One takes then 5--dimensional charged Dirac operator $\gamma^{\alpha}
\nabla_{\alpha}$ and considers spinors that are equivariant with
respect to the fifth coordinate $x^{5} = \phi: $
\be
{\partial \psi \over \partial \phi} = - i \psi \ . 
\ee
This first--order,  four--component spinor  (called L\'evy--Leblond
equation in Ref.  \cite{kun}) equation reduces then easily to the
second--order,  two--component Schr\"{o}\-din\-ger Pauli equation
with the correct Land\'e factor.\footnote{For an alternative
detailed derivation see \cite{can} }

We finish this section with pointing to the Ref.  \cite{jamo}, 
where Schr\"{o}din\-ger's quantization is discussed in details and
where a probabilistic interpretation of generally covariant
Schr\"{o}dinger equation is given using the bundle
$L^{2} (E_{t})$ of Hilbert spaces.  The parallel transport induced
by the quantum connection is shown in \cite{jamo} to be directly
related to Feynman amplitudes. 
\section{Coupled Quantum and Classical Systems}
\subsection{Preliminaries}
Replacing Schr\"odinger's evolution, which governs the dynamics
of pure states, by an equation of the
Liouville type, that describes time evolution of mixed states,
is a necessary step -- but it does not suffice for modeling
of {\sl real world events}.  One must take,  to this end,  two further steps. 
First of all we should admit that in our reasoning,  our communication,  our
description of {\sl facts} -- we are using classical logic.  Thus {\sl somewhere }\
in the final step of transmission of information from quantum systems to
macroscopic recording devices and further, to our senses and minds,
a {\sl translation}\ between quantum and classical
 should
take place.  That such a translation is necessary is evident also when we
consider the opposite direction:  to {\sl test} a physical theory we perform 
{\sl controlled experiments}.  But some of the {\sl controls} are always of
classical nature -- they are external parameters with concrete numerical
values.  So,  we need to consider systems with both quantum and classical
degrees of freedom,  and we need evolution equations that enable communication 
in both directions,  i.e. :  \begin{itemize}
\item {\sl flow of information from quantum to classical} \\and \item 
{\sl control of quantum states and processes by
classical parameters}\ . 
\end{itemize}

\subsection{Completely Positive Maps}
We begin with a brief recall of relevant mathematical concepts.  
Let ${\cal A}$ be a $C^{\star}$ -- algebra.  We shall always assume that $
{\cal A}$ has unit $I$.  An element $A\in {\cal A}$ is {\sl positive},  $A\geq 0$, 
iff it is of the form $B^\star B$ for some $B\in {\cal A}$.  Every element of
a $C^{\star}$--algebra is a linear combination of positive elements.  A
linear functional $\phi :  {\cal A} \rightarrow \complex$ is {\sl positive} iff 
$A\geq 0$
implies $\phi  (A) \geq 0$.  Every positive functional on a $C^{\star}$
--algebra is continuous and $\| \phi\| = \phi  (I). $ Positive functionals of
norm one are called {\sl states}.  The space of states is a convex set.  Its
extremal points are called {\sl pure states}.  The canonical
 {\sl GNS construction}
allows one to associate with each state $\omega$ a representation $
\pi_\omega $ of ${\cal A}$ on a Hilbert space ${\cal H}_\omega$,  and a
cyclic vector $\Psi_\omega\in{\cal H}_\omega$ such that $ (\Psi_\omega
, \pi_\omega  (A)\Psi_\omega ) = \omega  (A) , \,  A\in {\cal A} . $
Irreducibility of $\pi_\omega$ is then equivalent to purity of $\omega . $

Quantum theory gives us a powerful formal language and
statistical algorithms for
describing general physical systems.
Physical {\sl quantities}\ are coded there by
Hermitian elements of a $C^{\star }$--algebra ${\cal A}$ of {\sl observables}, 
while information about their values  (quantum algorithms deal,  in general, 
only with statistical information)  is coded in states of 
${\cal A}$ .  Pure states correspond to a maximal possible information.  For
each state $\omega , $ and for each $A=A^{\star }\in {\cal A}$ the  (real)
number
$\omega
 (A)$ is interpreted as {\sl expectation value} of observable $A$ in state $
\omega , $ while 
$$
\delta _\omega ^2 (A)\doteq \omega  ( (A-\omega  (A))^2)=\omega 
 (A^2)- (\omega  (A))^2 
$$
is the {\sl quadratic dispersion} of $A$ in the state $\omega . $ It is assumed
that repeated {\sl measurements} of $A$
\footnote{Our point is that "measurement" is an undefined concept in
standard quantum theory, and that the probabilistic interpretation
must be, because of that, brought from outside. What we propose
is to {\sl define}\, measurement as a CP semigroup coupling between a
classical and a quantum system and to {\sl derive} the probabilistic
interpretation of the quantum theory from that of the classical one.}
made on systems 
{\sl prepared in a state} $
\omega $ will give a sequence of values $a_1, \ldots , a_n$ so that 
{\sl approximately} ${\frac
1N}\sum_{i=1}^Na_i\approx \omega  (A), $ and ${\frac 1N}\sum  (a_i)^2- ({\frac 1N
}\sum a_i)^2\approx \delta _\omega ^2. $ If ${\cal A}$ is Abelian,  then it is
isomorphic to an algebra of functions ${\cal A}\approx C (X). $ Then pure
states of ${\cal A}$ are dispersion free -- they are parameterized by points $
x\in X$ and we have $\omega _x (A)=A (x). $ This corresponds to a {\sl classical
theory}:  all observables mutually commute and maximal possible information
is without any statistical dispersion.  In the extreme opposition to that is
{\sl pure quantum theory} -- here defined as that one 
when ${\cal A}$ is a factor, that is has a trivial centre.  The centre
${\cal Z (A)}$ of a $C^\star$--algebra ${\cal A}$ is defined as
${\cal Z (A)}=\{C\in {\cal A}
: AC=CA\, , \, A\in {\cal A}\}. $ In general $\complex \cdot I\subset {\cal Z (A)}
\subset 
{\cal A}. $ If {\cal Z (A) = A} -- we have pure classical theory.  If ${\cal 
Z (A)}=\complex \cdot I$ -- we have pure quantum theory.  In between we have 
a theory
with {\sl superselection rules}.  Many physicists believe that the ''good
theory'' should be a ''pure quantum'' theory.  But I know of {\sl no one good
 reason}\
why this should be the case.  In fact,  we will see that cases with a
nontrivial ${\cal Z (A)}$ are interesting ones.  Of course,  one can always
argue that whenever we have an algebra with a nontrivial centre -- it is a
subalgebra of an algebra with a trivial one,  for instance of $B ({\cal H})$
-- the algebra of all bounded operators on some Hilbert space.  This is,
however, not a good argument -- one could argue as well that we do not need
to consider different groups as most of them are subgroups of ${\cal U (H)}$
-- the unitary group of an infinite dimensional Hilbert space -- so why
to bother with others?

Let ${\cal A, B}$ be $C^{\star}$--algebras.  A linear map $\phi :  {\cal A}
\rightarrow {\cal B}$ is {\sl Hermitian} if $\phi  (A^{\star}) = \phi
 (A)^{\star }. $\ It is {\sl positive} iff $A\geq 0 , $ $A\in {\cal A}$ implies $
\phi  (A)\geq 0 . $ Because Hermitian elements of a $C^{\star}$--algebra are
differences of two positive ones -- each positive map is automatically
Hermitian.  Let ${\cal M}_n$ denote the $n$ by $n$ matrix algebra,  and let $
{\cal M}_n ({\cal A}) = {\cal M}_n \otimes {\cal A}$ be the 
algebra of $n\times n$ matrices with entries from ${\cal A}. $ Then 
$
{\cal M}_n ({\cal A})$ carries a natural structure of a $C^{\star}$--algebra.  
With respect to this structure a  matrix ${\bf 
A}= (A_{ij})$ from ${\cal M}_n  ({\cal A})$ is positive iff it is  a sum of
matrices of the form $ (A_{ij}) =  (A_i^{\star } A_j ), \,  $ $A_i\in {\cal A} . $ 
If ${\cal 
A}$ is an algebra of operators on a Hilbert space ${\cal H}$,  then ${\cal M}
_n ({\cal A})$ can be considered as acting on ${\cal H}^n \doteq {\cal H}
\otimes C^n = \oplus_{i=1}^n {\cal H} . $ Positivity of ${\bf A}= (A_{ij})$ is
then equivalent to $ ({\bf \Psi}, {\bf A} {\bf \Psi} )\geq 0\,  ,  \,  {\bf \Psi}
\in {\cal H}^n ,  $ or equivalently,  to $\sum_{i, j}  (\Psi_i ,  A_{ij} \Psi_j
) \geq 0 $ for all $\Psi_1, \ldots , \Psi_n \in {\cal H} . $ \\
A positive map $
\phi$ is said to be {\sl completely positive} or,  briefly,   {\sl CP}
 iff $\phi \otimes id_n
: {\cal A }\otimes {\cal M}_n \rightarrow {\cal B}\otimes {\cal M}_n $ defined
by $  (\phi\otimes id_n)  (A\otimes M ) = \phi  (A)\otimes M , \,  M\in {\cal M}_n
$, is positive for all $n=2, 3, \ldots . $ When written explicitly,  complete
positivity is equivalent to \be \sum_{i, j=1}^n B_i^{\star}\phi
 (A_i^{\star}A_j)B_j \geq 0 \ee for every $A_1, \ldots , A_n \in {\cal A}$ and
$B_1, \ldots , B_n \in {\cal B} . $ In particular every homomorphism of 
$C^{\star}$ algebras is completely positive.  One can also show that if either 
${\cal A}$ or ${\cal B}$ is Abelian,  then positivity implies complete
positivity.  Another important example:  if ${\cal A}$ is a $C^{\star }$ 
algebra of
operators on a Hilbert space ${\cal H}$,  and if $V\in {\cal B} ({\cal H}) , $
then $\phi  (A) = VAV^{\star}$ is a CP map $\phi :  {\cal A}
\rightarrow \phi ({\cal A}) . $ The celebrated Stinespring theorem gives us a
general form of a CP map.  Stinespring's construction can be
described as follows.  Let $\phi :  {\cal A}\rightarrow {\cal B}$ be a
CP map.  Let us restrict to a case when $\phi  (I)=I . $ 
Let ${\cal B}$ be
realized as a norm closed algebra of bounded operators on a Hilbert space 
${\cal H}$.  One takes then the algebraic tensor product ${\cal A}\otimes 
{\cal H}$ and defines on this space a sesquilinear form $ <\,  , \,  >$ by 
\be <A\otimes \Psi ,  A^{\prime}\otimes
\Psi^{\prime}>= (\Psi ,  \phi  (A^{\star}A^{\prime} ) \Psi^{\prime}) .  \ee
This scalar product is then positive semi--definite because of complete
positivity of $\phi . $ Indeed,  we have \be <\sum_i A_i\otimes\Psi_i ,  \sum_j
A_j\otimes\Psi_j > = \sum_{i, j}  (\Psi_i, \phi  (A_i^{\star } A_j ) \Psi_j )
\geq 0 .  \ee  Let ${\cal N}$ denote the kernel of $ <\,  , \,  >$ .  
Then ${\cal A}\otimes {\cal H} / {\cal N} $ is a
pre--Hilbert space.  One defines a representation $\pi$ of ${\cal A}$ on 
${\cal A}\otimes {\cal H}$ by $\pi  (A) : \,  A^{\prime}\otimes\Psi\longmapsto
AA^{\prime}\otimes\Psi . $ One shows then that ${\cal N}$ is invariant under $
\pi  ({\cal A})$,  so that $\pi $ goes to the quotient space.  Similarly,  the
map ${\cal H}\ni\Psi \mapsto I\otimes \Psi \in {\cal A}\otimes{\cal H}$ 
defines an isometry $V :  {\cal H}
\rightarrow {\cal A}\otimes {\cal H} / {\cal N} . $ We get then $
\phi  (A) = V^{\star}\pi  (A) V $ on the completion ${\cal H}_{\phi } $ of $
{\cal A}\otimes {\cal H} / {\cal N} . $ \vspace{0.5cm}\\
\begin{theorem}[Stinespring's Theorem]
Let ${\cal A}$ be a
$C^{\star}$--algebra with unit and let $\phi : {\cal A}\rightarrow
{\cal B} ({\cal H})$ be a CP map.  Then there exists a
Hilbert space ${\cal H}_\phi , $ a representation $\pi_\phi$ of ${\cal A}$
on ${\cal H}_\phi ,  $ and a bounded linear map $V :  {\cal
H}\rightarrow {\cal H}_\phi $ such that  
\be \phi  (A) = V^{\star}\pi_\phi
 (A) V .  
\ee $V$ is an isometry iff $\phi$ is unital i.e.  iff $\phi$
maps the unit of ${\cal A}$ into the identity operator of ${\cal H}. $
If ${\cal A}$ and ${\cal H}$ are separable,  then ${\cal H}_\phi $ can
be taken separable. 
\end{theorem}
\vspace{1cm}
The space of CP maps from ${\cal A}$ to ${\cal B} ({\cal H})$
is a convex set.  Arveson \cite{arv} proved that $\phi $ is an extremal
element of this set iff the representation $\pi_\phi$ above is irreducible. 

\subsection{Dynamical Semigroups}

A {\sl dynamical semigroup} on a $C^{\star }$--algebra of operators ${\cal A}$ 
is
a strongly continuous semigroup of CP maps of {\cal A} into
itself.  A semigroup $\alpha _t , $ is norm continuous 
iff its infinitesimal
generator $L$ is bounded as a linear map $L: {\cal A}\rightarrow {\cal A} . $
We then have \be
\alpha _t=\exp (t L) \thinspace , \thinspace t\geq 0 .  \ee
The right hand side is,  in this case,   a norm convergent series {\sl for all}\
real values of $t, $ however for $t$ negative the maps $\exp (tL): {\cal A}
\rightarrow {\cal A}$, although Hermitian, need not be positive.  \\ 
Evolution of observables
gives rise,  by duality,  to evolution of positive functionals.  One defines $
\alpha ^t (\phi ) (A)=\phi  (\alpha _t (A)). $ Then $\alpha _t$ preserves the
unit of ${\cal A}$ iff $\alpha ^t$ preserves normalization of states.  A
general form of a generator of a dynamical semigroup in finite dimensional
Hilbert space has been derived by Gorini,  Kossakowski and Sudarshan \cite
{koss},  and Lindblad \cite{lin} gave a general form of a bounded generator of
a dynamical semigroup acting on the algebra of all bounded operators ${\sl B}
 ({\cal H}). $ It is worthwhile to cite, after Lindblad, his
 original motivation:

"The dynamics of a finite closed quantum system is conventionally
represented by a one--parameter group of unitary transformations in Hilbert
space.  This formalism makes it difficult to describe irreversible processes
like the decay of unstable particles,  approach to thermodynamic equilibrium
and measurement processes [$\ldots $].  It seems that the only possibility of
introducing an irreversible behaviour in a finite system is to avoid the
unitary time development altogether by considering non--Hamiltonian systems. "

In a recent series of papers \cite{bla1,bla2,bla3,bla4} Ph. Blanchard and
the present author were forced to introduce dynamical semigroups because of
another difficulty,  namely because of  impossibility of obtaining 
a nontrivial Hamiltonian
coupling of classical and quantum degrees of freedom in a system described
by an algebra with a non--trivial centre.  We felt that lack of a
dynamical understanding of quantum mechanical probabilistic postulates is
more than annoying.  We also believed that the word ''measurement'' instead 
of being
banned,  as suggested by J.  Bell \cite{bel1,bel2},  can be perhaps given a precise
and acceptable meaning.  We suggested that {\sl a measurement process is a 
coupling of a
quantum and of a classical system,  where information about quantum state is
transmitted to the classical recording device by a dynamical semigroup of
the total system\ }.  It is instructive to see that such a transfer of
information can not indeed be accomplished by a Hamiltonian or,  more generally, 
 by
any automorphic evolution\footnote{For a discussion of this fact in a broader 
context
of algebraic theory of superselection sectors -- cf. 
Landsman \cite[Sec. 4. 4]{lan}. Cf. also the no--go result by 
Ozawa \cite{oza}}.  To this end consider an algebra ${\cal A}$ with
centre ${\cal Z}. $ Then ${\cal Z}$ describes classical degrees freedom.  Let 
$\omega $ be a state of ${\cal A}, $ then ${\omega}|_Z$ denotes its
restriction to ${\cal Z}. $ Let $\alpha _t$ be an automorphic evolution of $
{\cal A}, $ and denote $\omega _t=\alpha ^t (\omega ). $ Each $\alpha _t$ is
an automorphism of the algebra ${\cal A} , $  and so it leaves its centre 
invariant:  
$\alpha _t: {\cal Z}\rightarrow {\cal Z}. $ The crucial observation is that,
because of that fact,  the restriction ${\omega _t}|_{{\cal Z}}$ depends only 
on ${
\omega _0}|_{{\cal Z}}, $ as the evolution of states of ${\cal Z}$ is dual to
the evolution of the observables in ${\cal Z}. $ This shows that information
transfer from the total algebra ${\cal A}$ to its centre ${\cal Z}$ is
impossible -- unless we use more general,  non--automorphic evolutions. \\
>From the above reasoning it may be seen that the Schr\"odinger picture,  when
time evolution is applied to states,  is better adapted to a discussion of
information transfer between different systems.  The main properties that a
dynamical semigroup $\alpha ^t$ describing time evolution of states should
have are:  $\alpha ^t$ should preserve convex combinations,  positivity and
normalization.  One can demand even more -- it is reasonable to demand a
special kind of stability:  
that it should be always possible to extend the
system and its evolution in a trivial way,  by adding extra degrees of
freedom that do not couple to our system.\footnote{That requirement 
is also necessary to guarantee physical consistency of the whole
framework, as we always neglect some degrees of freedom as either
irrelevant or yet unknown to us.}  
That is exactly what is assured by 
{\sl complete positivity} of the maps $\alpha_t. $ One could also think that
we should require even more,  namely that $\alpha ^t$ transforms pure states 
into
pure states.  But to assume that would be already too much,
as one can prove that then
$\alpha^t $ must be dual to an automorphic evolution.  
It appears that information 
gain in one respect  (i.e. learning about the actual 
state of the quantum system) 
must be accompanied by information loss in another one -- as
going from pure states to mixtures implies entropy growth. 

We will apply the theory of dynamical semigroup to algebras with a non--trivial
centre.  In all our examples we will deal with tensor products of 
${\cal B} ({\cal H})$ and an Abelian algebra of functions.  
The following theorem by Christensen and Evans \cite{chr} generalizes the
results of Gorini,  Kossakowski and Sudarshan and of Lindblad to the case of
arbitrary $C^{\star }$--algebra. 

\begin{theorem}[Christensen -- Evans]
Let $\alpha_t = \exp  (L t)$ be a
norm--continuous semigroup of CP maps of a
$C^{\star }$-- algebra of operators ${\cal A}\subset {\cal B} ({\cal
H}) . $ Then there exists a CP map $\phi$
of ${\cal A}$ into the ultraweak closure ${\bar {\cal A}}$
and an operator $K\in {\bar {\cal A}}$ such that the
generator $L$ is of the form: 
\be
L (A) = \phi  (A) + K^{\star }A + AK \,  . 
\ee
\end{theorem}
We will apply this theorem to the cases of ${\cal A}$ being a von Neumann
algebra,  and the maps $\alpha_t$ being {\sl normal}.  Then $\phi$ can be
also taken normal.  We also have ${\bar {{\cal A}}} = {\cal A} , $
so that $K\in {\cal A} . $ We will always assume that $\alpha_t  (I) = I $ or, 
equivalently,  that $L (I)=0 . $  Moreover,  it is convenient to introduce $
H=i (K-K^{\star })/2 \in {\cal A}, $ then from $L (I)=0$ we get $K+K^{\star
}=-\phi  (I) , $ and so $K=-iH-\phi  (1)/2 . $ Therefore we have \be
L (A) = i\left[H, A\right]+\phi  (A) -\{ \phi  (1) , A\}/2 ,  \ee         
where $\{\,  ,  \,  \}$ denotes anticommutator.  Of particular interest to us
will be generators $L$ for which $\phi$ is extremal \footnote{It should 
be noticed, 
however,  that splitting of $L$ into $\phi$ and $K$ is,  in general,  not
unique -- cf.  e. g.  Refs \cite{dav} and \cite[Ch. III. 29--30]{par}. }.  
By the already
mentioned result of Arveson \cite{arv} this is the case when $\phi$ is
of the form \be
\phi  (A)=V^{\star }\pi  (A) V \,  ,  \ee   
where $\pi$ is an irreducible representation of ${\cal A}$ on a Hilbert
space ${\cal K} , $ and $V: {\cal H}\rightarrow {\cal K}$ is a bounded
operator  (it must be,  however,  such that $V^{\star}{\cal A} V \subset 
{\cal A
}$).  
\subsection{Coupling of Classical and Quantum Systems}
We consider a model describing a coupling between a quantum and
a classical system.  To concentrate on main ideas rather than on
technical details let us assume that the quantum system is described in
an $n$--dimensional Hilbert space ${\cal H}_q , $ and that it has as its
algebra of observables ${\cal B} ({\cal H}_q) \approx {\cal M}_n . $ Similarly, 
let us assume that the classical system has only a finite number of pure
states ${\cal S}=\{ s_1, \ldots , s_m\} . $ Its algebra of observables
${\cal A}_{cl}$ is then isomorphic to $\complex^m . $ For the algebra
of the total system we take
${\cal A}_{tot}={\cal A}_q \otimes {\cal A}_{cl}$ which is isomorphic to the
diagonal subalgebra of ${\cal M}_m  ({\cal A}_q) . $ Observables of the total
system are block diagonal matrices: 
$${\bf A}=diag (A_{\alpha}) = \pmatrix{A_1&0&0&\ldots&0\cr
0&A_2&0&\ldots&0\cr
&&\ldots&&\cr
0&0&0&\ldots&A_m\cr}, $$
where $A_{\alpha},   (\alpha=1, \ldots , m)$ are operators in ${\cal H}_q
. $
\footnote{It is useful to have the algebra  ${\cal A}_{tot}$
represented in such a form, as it enables us to apply the
theorem of Christensen--Evans.}
Both ${\cal A}_q$ and ${\cal A}_{cl}$ can be considered as
subalgebras  of ${\cal A}_{tot}$ consisting respectively of matrices
of the form
$diag (A, \ldots, A)\,  ,  A\in {\cal A}_q$ and $diag (\lambda_1
I_n, \ldots, \lambda_m I_n)\,  ,  \lambda_{\alpha}\in\complex . $ 
States of the quantum system are represented by positive,  trace one,  
operators on
${\cal B} ({\cal H}_q) . $ States of the classical system are
$m$--tuples of non--negative numbers $p_1, \ldots, p_m , $  with
$\sum_{\alpha} p_{\alpha}=1. $ States of the total system are
represented by block diagonal matrices
$ \rho =diag (\rho_1, \ldots, \rho_m) , $ with ${\cal B} ({\cal
H}_q)\ni \rho_{\alpha}\geq 0$ and $\sum_{\alpha} Tr (\rho_{\alpha})=1
. $ For the expectation value we have  $\rho  ({\bf A}) =\sum_{\alpha}
Tr (\rho_\alpha A_\alpha ) . $ Given a state $\rho$ of the total
system,  we can trace over the quantum system to get an effective state
of the classical system
$p_\alpha =Tr (\rho_\alpha ) ,  $ or we can trace over the classical
system to get the effective state of the quantum system
${\hat\rho}=\sum_\alpha\rho_\alpha . $
\footnote{One can easily imagine a more general situation when
tracing over the classical system will not be meaningful.  This can
happen if we deal with several {\sl phases} of the quantum system, 
parameterized by the classical parameter $\alpha$.  It may then happen
that the total algebra is not the  tensor product algebra.  For
instance,  instead of one Hilbert space ${\cal H}_q , $ we may have, 
for each value of $\alpha , $ a Hilbert space ${\cal H}_{q, \alpha}$ 
of dimension $n_\alpha$ . }\\
Let us consider dynamics.  Since the classical system has a discrete
set of pure states,  there is no non--trivial and continuous  time
evolution for the
classical system that would map pure states into pure states. 
 As for the quantum system,  we can have a Hamiltonian dynamics,  with
the Hamiltonian possibly dependent on time and on the state of the 
classical
system $H (t) = diag (H (\alpha, t)) . $ As we already know a non--trivial
coupling between both systems is impossible without a dissipative
term,  and the simplest dissipative coupling is of the form
$L ({\bf A})=V\pi ({\bf A})V^\star , $ where $\pi$ is an irreducible
representation of the algebra ${\cal A}_{tot}$ in a Hilbert space
${\cal H}_\pi , $ and $V: {\cal H}_q\rightarrow{\cal H}_\pi$ is a linear
map.  It is easy to see that such an $L (A)$ is necessarily of the form: 
$$L ({\bf A})={\bf V}^\star {\bf A} {\bf V} , $$
where ${\bf V}$ is an $m\times m$ block matrix with only one non--zero
entry.  A more general CP map of ${\cal A}_{tot}$ is of the same form, 
but with ${\bf V}$ having at most one non--zero element in each of its
rows. \\
Let us now discuss   desired couplings in somewhat vague,  but more
intuitive,  physical terms.  We would like to write down a coupling that
enables transfer of information from quantum to classical system. 
There may be many ways of achieving this aim -- the subject is new and
there is no ready theory that fits.
We will see however that a naive
description of a coupling works quite well in many cases.  The idea is 
that the simplest coupling {\sl associates} to a {\sl property} of the
quantum system a {\sl transformation} of the actual state of the
classical system.  \\
{\sl Properties} are,  in quantum theory,  represented by projection
operators.  Sometimes one considers also more general,  {\sl unsharp} or
{\sl fuzzy } properties.  They are represented by positive elements of
the algebra which are bounded by the unit.  A {\sl measurement} should
discriminate between mutually exclusive and exhaustive properties. 
Thus one usually considers a family of mutually orthogonal projections
$e_i$ of sum one.  With an {\sl unsharp measurement} one associates 
a family
of positive elements $a_i$ of sum one. \\ 
As there is no yet a complete,  general,  theory
of dissipative couplings of classical and quantum systems,  the best we
can do is to show some characteristic examples. It will be done in the
following section.  For every example a piecewise deterministic random
process  will be described that takes place on the space of pure
states of the total system
\footnote{One may wonder what does that mean mathematically, as
the space of pure states of a $C^{\star}$ algebra is, from
measure--theoretical point of view, a rather
unpleasant object. The answer is that the only measures on the
space of pure states of the quantum algebra will be the Dirac
measures.}
 and which reproduces the Liouville
evolution of the total system by averaging over the process.
A theory of piecewise
deterministic  (PD) processes is described in a recent book by M. H. 
Davis \cite{dav}.  Processes of that type,  but without a non--trivial
evolution of the classical system,  were discussed also in physical
literature -- cf.  Refs \cite{car,dal,dum,gar}.
\footnote{Thanks are due to N.  Gisin for pointing out these
references. }
We will consider Liouville equations of the form
\be {\dot \rho} (t) = -i[H , \rho  (t)] + {\sum}_{i} \left  ( V_i
\rho  (t) V_i^{\star} - {1 \over 2}
\{V_i^{\star} V_i ,  \rho  (t)\}
\right ) ,  \label{eq: lio}\ee  where in  general $H$ and the $V_i$ can
explicitly depend on time. 
The $V_i$ will be chosen as tensor products
$V_i=\sqrt{\kappa} e_i\otimes\phi_i$,  where
$\phi_i$ act as transformations
\footnote{Or, more precisely, as Frobenius--Perron operators. Cf.
Ref.\cite{las} for definition and examples of Frobenius--Perron and
dual to them Koopman operators.} 
 on classical  (pure) states. 
\subsection{Examples of Classical--Quantum Couplings}
\subsubsection{The Simplest  Coupling}
First,  we consider only one orthogonal projector $e$  on the
two--dimensio\-nal Hilbert space
${\cal H}_q=\complex^2. $
To define the dynamics we choose the coupling operator $V$ in the
following way: 
\be V = \sqrt{\kappa} \pmatrix{0,     & e \cr
                   e, & 0     } . \ee The Liouville equation
 (\ref{eq: lio} ) for the density matrix $\rho = diag (\rho_1, \rho_2)$ of
the total system reads now
\be
\ba{rl}
\ds{{\dot \rho}_1 = }\! &\ds{-i[H, \rho_1]+\kappa  (e\rho_2e-{1\over
2}\{e, \rho_1\}}), \cr\cr 
\ds{{\dot \rho}_2 = }\! &\ds{-i[H, \rho_2]+\kappa  (e\rho_1e-{1\over
2}\{e, \rho_2\}). }\cr\cr
\ea
\ee For this particularly simple  coupling the effective quantum state
$\hat{\rho}={\pi}_q (\rho )=\rho_1+\rho_2$ evolves independently of the
state of the classical system.  One can say that here we have only
transport of information from the quantum system to the classical one. 
We have: 
\be   {\dot {\hat\rho }} = -i[H, {\hat\rho}]+\kappa
 (e{\hat\rho}e-{1\over 2}\{e, {\hat\rho}\}). 
\ee
The Liouville equation  (\ref{eq: lio} ) describes time evolution of
statistical states of the total system. \\ 
Let us describe now a the PD process associated to this equation. 
Let
$T_t$ be a one-parameter semigroup of  (non-linear) transformations of
rays in
$\complex^2$ given by
\be T (t)\phi  ={\phi  (t) \over \Vert \phi  (t) \Vert}, 
\ee where
\be
\phi  (t)=\exp\left ({-iHt-{\kappa\over 2}et}\right)
\phi . 
\ee    Suppose we start with the quantum system in a pure state
$\phi_0$,  and  the classical system in a state
$s_1$  (resp.  $s_2$).  Then $\phi_0$ starts to evolve according to the
deterministic  (but non--linear Schr\"odinger) evolution $T (t)\phi_0 $ until a
jump occurs at time $t_1$.  The time $t_1$ of the jump is governed by
an inhomogeneous Poisson process with the rate function
$\lambda  (t) = \kappa \Vert e T (t)\phi_0 \Vert^2$.  Classical system
switches from $s_1$ to $s_2$  (resp.  from $s_2$ to $s_1$),  while  
$T (t_1)\phi_0$ jumps to
$\phi_1=eT (t_1)\phi_0/\Vert eT (t_1)\phi_0 \Vert$,  
and the process starts again.  With the initial
state being an eigenstate of $e$,  $e\phi_0=\phi_0$,  the rate
function $\lambda$ is approximately constant and equal to $\kappa$. 
Thus
${1/\kappa}$ can be interpreted as the expected time interval between
the successive jumps.  \\
More details about this model illustrating the quantum Zeno effect
can be found in Ref.  \cite{bla2}. 
\subsubsection{Simultaneous "Measurement" of Several Noncommuting Observables}

Using somewhat pictorial language we can say that in the previous
example each {\sl actualization} of the property
 $e$ was causing a flip in the classical system.  In the present
example,  which is a non--commutative and fuzzy generalization of the model 
discussed in \cite{bla1},  we
consider
$n , $   in general {\sl fuzzy},  properties
$a_i=a_i^\star , \,  i=1, \ldots , n. $ 
The Hilbert space ${\cal H}_q$ can be completely arbitrary, 
for instance $2$--dimensional.    We will
denote
$a_0^2\doteq\sum_{i=1}^n a_i^2 . $ The 
$a_i$--s need not be projections,  and the different $a_i$--s 
need not to commute.   The classical system
is assumed to have
$n+1$ states
$s_0, s_1, \ldots , s_n$ with $s_0$ thought of as an initial,  
neutral state. 
  To
each actualization of the property $a_i$ there will be associated a
flip between
$s_0$ and
$s_i . $  Otherwise
the state of the classical system will be unchanged.  To this end we
take 
$$V_{1} ={\sqrt\kappa}
 \pmatrix{0,     & a_{1} ,  &0 ,  &\ldots ,  &0 ,    
&0
\cr
                   a_{1}, & 0     ,  &0 ,  &\ldots ,  &0 ,     &0 \cr
                   0,    & 0     ,  &0 ,  &\ldots ,  &0 ,     &0\cr
\ldots&\ldots   &\ldots&\ldots &\ldots&\ldots\cr
                   0    , & 0     ,  &0 ,  &\ldots ,  &0 ,    &0} , $$

$$V_{2} = {\sqrt\kappa}\pmatrix{0,     & 0     ,  & a_{2} ,  &\ldots ,  &0
,  &0\cr
                   0    , & 0     ,  &0 ,  &\ldots ,  &0 ,  &0\cr
                   a_{2}, & 0     ,  &0 ,  &\ldots ,  &0 ,  &0\cr
                   0    , & 0     ,  &0 ,  &\ldots ,  &0 ,  &0\cr
                   \ldots&\ldots   &\ldots&\ldots&\ldots&\ldots\cr
                   0    , & 0     ,  &0 ,  &\ldots ,  &0 ,  &0} , $$
\vspace{0.5cm} $$\ldots$$ \vspace{0.5cm}
$$V_{n} = {\sqrt\kappa}\pmatrix{0,     & 0     ,  & 0 ,  &\ldots ,  &0 , 
&a_{n}\cr
                   0    , & 0     ,  &0 ,  &\ldots ,  &0 ,  &0\cr
                   \ldots&\ldots   &\ldots&\ldots&\ldots&\ldots\cr
                   0    , & 0     ,  &0 ,  &\ldots ,  &0 ,  &0\cr
                   a_{n}    , & 0     ,  &0 ,  &\ldots ,  &0 ,  &0} . $$

The Liouville equation takes now the following form: 
\be {\dot \rho_0}= -i[H, \rho_0]+
\kappa \sum_{i=1}^n a_i\rho_i a_i-{\kappa\over 2} \{ a_0^2, \rho_0\} , 
\ee
\be {\dot \rho_i}=-i[H, \rho_i]+
\kappa a_i\rho_0 a_i -{\kappa\over 2}\{ a_i^2, \rho_i\} . \ee 
We will derive the PD process for this example in some more details, 
so that a general method can be seen.  First of all we transpose the
Liouville equation so as to get time evolution of observables; we
use the formula 
\be
\sum_\alpha Tr ({\dot A_\alpha}\rho_\alpha ) = \sum_\alpha
Tr (A_\alpha{\dot \rho_\alpha}) . 
\ee
In the particular case at hand the evolution equation for observables
looks almost exactly the same as that for states: 
\be {\dot A_0}= i[H, A_0]+
\kappa \sum_{i=1}^n a_iA_i a_i-{\kappa\over 2} \{ a_0^2, A_0\} , 
\ee
\be {\dot A_i}=i[H, A_i]+
\kappa\,  a_iA_0 a_i -{\kappa\over 2}\{ a_i^2, A_i\} . \ee  
Each observable ${\bf A}$ of the total system defines now a function
$f_{\bf A} (\psi, \alpha )$ on the space of pure states of the total
system
\be 
f_{\bf A} (\psi, \alpha ) =  (\psi , A_\alpha \psi ) . 
\ee
We have to rewrite the evolution equation for observables in terms of
the functions $f_{\bf A} . $  To this end we compute the
expressions $ (\psi , {\dot A}_\alpha \psi) . $
Let us first introduce the Hamiltonian vector field $X_H$ on the
manifold of pure states of the total system: 
\be
 (X_H f) (\psi, \alpha)={d\over dt}f (e^{-iHt}\psi )\vert_{t=0}. 
\ee
Then the terms $ (\psi ,  i[H, A_\alpha ] \psi )$ can be written as
$ (X_H f_{\bf A} ) (\psi, \alpha) . $
We also introduce vector field $X_D$ corresponding to non--linear
evolution: 
\be
 (X_D f) (\psi, \alpha ) = {d\over dt}f
\left ( {
\exp  ({-\kappa t a_\alpha^2 }/2)\psi
\over
\| {\exp  ({-\kappa t a_\alpha^2}/2)\psi\|}}\right) \vert_{t=0} . \ee
Then evolution equation for observables can be written in
a Davis form:  
\be
\begin{array}{ll}
\displaystyle{
{d\over dt}f_{\bf A}  (\psi , \alpha ) 
} & =   \displaystyle{ ( (X_H+X_D)f) (\psi , \alpha) \  + }
\cr\cr
 & + \  \displaystyle{\lambda  (\psi , 
\alpha)\sum_\beta\int Q (\psi, \alpha ;d\phi, \beta ) \left (f_{\bf A}
 (\phi, \beta)-f_{\bf A} (\psi, \alpha)\right) , 
}
\end{array}
\ee
where $Q$ is a matrix of measures,  whose non--zero entries are: 
\be
Q (\psi, 0 ;d\phi, i) = {\| a_i\psi\|^2\over \| a_0\psi\|^2  }\delta
\left (
\phi - {a_i\psi \over \| a_i\psi\| }\right)d\phi ,  
\ee
\be
Q (\psi, i ;d\phi, 0) = \delta \left (
\phi - {a_i\psi \over \| a_i\psi\| }\right)d\phi, 
\ee
while
\be
\lambda  (\psi , \alpha )=\kappa \| a_\alpha\psi\|^2 . 
\ee 
The symbol $\delta \left (\phi - \psi\right)d\phi$ denotes here the
Dirac measure concentrated at $\psi . $\\
We describe now PD process associated to the above semigroup. 
There are $n$ one-parameter  (non-linear) semigroups $T_\alpha  (s)$ acting on the 
space  of pure states of the quantum system via
$$\psi \mapsto T_\alpha  (t) \psi ={ W_\alpha  (t) \psi  \over
\| W_\alpha  (t)\psi \|}, $$ where
$$W_\alpha  (t)=\exp[-iHt-{\kappa\over 2} a_\alpha^2 t]. $$ If initially the
classical system is in a pure state $\alpha$,   and quantum system  in
a pure state
$\psi$,   then quantum system evolves deterministically according to the 
semigroup
$T_\alpha$:\, $\psi (t)\mapsto T_\alpha  (t)\psi$. 
 The classical system then jumps at
the time instant
$t_1$,  determined by  the inhomogeneous Poisson process with rate function
$\lambda_\alpha = \lambda(\psi,\alpha)$.
If the classical system was in one of the states
$j=1, 2, \ldots , n$,  then it jumps to
$0$ with probability one,  the quantum state jumps at the same time to the state
$a_j\psi  (t_1) / \| a_j\psi  (t_1)\|$. 
If,  on the other hand,  it was in the state $0$,  then it jumps to one of the
states $j$ with  probability
$\|a_j\psi  (t_1)\|^2/\|a_0 \psi  (t_1)\|^2$.  The quantum state jumps at the same
time to
$a_j\psi (t_1)/\|a_j\psi (t_1)\|. $ Let 
$$F_\alpha  (t)=\exp[-\int_0^t \lambda_\alpha  (T_\alpha  (s)\psi)ds]. $$
 
Then $F_\alpha$ is the distribution of $t_1$ - the first jump time.  More
precisely, 
$F_\alpha  (t)$ is the survival function for the state $\alpha$:
$$F_\alpha  (t) = P[t_1> t]. $$ Thus the probability distribution of the jump is
$p (t)=-dF_\alpha  (t)/dt ,  $ and the expected jump time is $\int_0^{+\infty } t\, 
p (t) dt. $  The probability that the jump will occur between $t$ and
$t+dt$, {\sl provided it did not occur yet}/, is
equal to
$1-\exp\left(\int_t^{t+dt}\lambda_{\alpha}(s)ds\right)\approx
\lambda_{\alpha}(t)dt$.
Notice that this depends on the actual state $(\psi,\alpha)$.
However,  as numerical
computation show,  the dependence  is negligible and approximately
 jumps occur always after time $t_1=1/\kappa$. 
\footnote{This sequence of transformation on the space of pure states of the
quantum system can be thought of as a nonlinear version of Barnsley's Iterated
Function System  (cf.  e. g.  \cite{bar}  }
\subsubsection{Coupling to All One--Dimensional Projections}

In the previous example the coupling between classical and quantum
systems involved a finite set of non--commuting observables.  In the
present one we will go to the extreme -- we will use {\underline all}
one--dimensional projections in the coupling.  One can naturally discover
such a model when looking for a precise answer to the question: 
\begin{center} {\sl how
to determine state of an individual quantum system ?}
\end{center}
For some time I was sharing the predominant opinion
 that a positive answer to this question can
not be given,  as there is no observable to be measured that answers the
question:  {\sl what state our system is in ?}.  Recently Aharonov and
Vaidman
\cite{aha} discussed this problem in some details.  \footnote{I do not
think that they found the answer,  as their arguments 
are circular,  and they seem to be well aware of this circularity. } 
The
difficulty here is in the fact that we have to discriminate between
non--orthogonal projections  (because {\sl different} states are not
necessarily {\sl orthogonal}),  and this implies necessity of 
simultaneous measuring
of non--commuting observables.  There have been many papers discussing
such measurements,  different authors taking often different positions.  
However they all seem to agree on  the fact that {\sl
predictions} from such measurements are necessarily {\sl fuzzy}.  
This fuzziness
being directly related to the Heisenberg uncertainty relation for
non--commuting observables.  Using methods and ideas presented 
in the previous 
sections of this chapter it is possible to build
models corresponding to the intuitive idea of a simultaneous
measurement of several non--commuting observables,  like,  for instance, 
 different spin components,  
positions \underline{and}
momenta etc.  A simple example of such a model was given in the 
previous section.  
After playing for a while with similar models it is
natural to think of a coupling between a quantum system and a
classical device that will result in a determination of the quantum state
by the classical device.  Ideally,  after the interaction,  the classical
"pointer" should point at some vector in a model
 Hilbert space.  This vector should
represent  (perhaps,  with some uncertainty) the actual state of the
quantum system.  The model that came out of this simple idea,  and which
we will now discuss,  does not achieve this goal.  But it is instructive, 
as it shows that models of this kind are possible.  I believe that one day
somebody will invent a better model,  a model that can be proven to be
optimal,  giving the best determination with the least disturbance.  Then
we will learn something important about the nature of quantum states. 

Our model will be formulated for a $2$--state quantum system.  It is
rather straightforward to rewrite it for an arbitrary $n$-state system, 
but for $n=2$ we can be helped by our visual imagination.  Thus we take
${\cal H}_q=\complex^2$ for the Hilbert space of our quantum system.  We
can think of it as pure spin $1/2$.  Pure states of the system form up
the manifold ${\cal S}_q\equiv\complex P^2$ which is isomorphic to the
$2$-sphere $S^2=\{ {\bf n}\in {\bf R}^3: {\bf n}^2=1\}$. 
Let ${\bf \sigma}=\{\sigma_i\}$, 
$i=1, 2, 3$ denote the Pauli $\sigma$--matrices.  Then for each ${\bf n}\in S^2$
the operator $\sigma ({\bf n})={\bf\sigma\cdot n}$ has eigenvalues
$\{+1, -1\} $.  We denote by $e ({\bf n})= (I+\sigma ({\bf n}))/2$ the projection
onto the $+1$--eigenspace.  \\
For the space ${\cal S}_{cl}$ of pure states of the classical system we
take also $S^2$ -- a copy of ${\cal S}_q$.  Notice that
$S^2$ is a homogeneous space for $U (2)$.  Let $\mu$ be the $U (2)$
invariant measure on $S^2$ normalized to $\mu  (S^2)=1$.  In spherical
coordinates we have $d\mu=sin (\theta)d\phi d\theta /4\pi$. 
We denote ${\cal H}_{tot}=L^2 ({\cal S}_{cl}, {\cal H}_q, d\mu )$ the 
Hilbert space of the total system,  and by 
${\cal A}_{tot}=L^{\infty} ({\cal S}_{cl}, 
{\cal L} ({\cal H}_q), d\mu )$ its von Neumann algebra of observables. 
 Normal states of ${\cal A}_{tot}$ are 
of the form $$\rho: {\bf A}\mapsto\int Tr (A ({\bf n})\rho ({\bf n}))d\mu ({\bf n}), $$
where
$\rho\in L^{\infty} ({\cal S}_{cl},  {\cal L} ({\cal H}_q), d\mu )$  satisfies
$$\rho  ({\bf n})\geq 0 , \,  {\bf n}\in {\cal S}_{cl} , $$
$$\int Tr \left (\rho  ({\bf n})\right) d\mu  ({\bf n}) =1 . $$
We proceed now to define the coupling of the two systems.  There will be two
constants:  
\begin{itemize}
\item $\kappa$ -- regulating the time rate of jumps 
\item $\omega$ -- entering the quantum Hamiltonian 
\end{itemize}
 The idea is that if the quantum system is at some pure state
${\bf n}_q$,  and if the classical system is in some pure states ${\bf n}_{cl}$, 
then
${\bf n}_{cl}$ will cause 
the Hamiltonian rotation of ${\bf n}_{q}$  around
${\bf n}_{cl}$ with frequency $\omega$,  while ${\bf n}_q$ will cause,  
after a random waiting time $t_1$ proportional to $1/\kappa$,  
a jump,  along geodesics,  to the "other side" of ${\bf n}_q$. 
The classical transformation involved 
is nothing but a geodesic symmetry on the symmetric
space $\complex P^2=U (2)/ (U (1)\times U (1))$. 
It has
the advantage that it is a measure preserving transformation.  It has
a disadvantage because ${\bf n}_{cl}$ {\sl overjumps} ${\bf n}_q$.  \\
We will
use the notation ${\bf n} ({\bf n}')$ to denote the $\pi$ rotation of 
${\bf n}'$ around ${\bf n}$.  Explicitly: 
$${\bf n} ({\bf n}')=2 ({\bf n\cdot n}'){\bf n}-{\bf n}' . $$
For each ${\bf n}$ we define $V_{\bf n}\in {\cal L} ({\cal H}_{tot})$
by
\be
\left ( V_{\bf n}\Psi\right)
 ({\bf n}')=\sqrt{\kappa} e ({\bf n})\Psi ({\bf n} (
{\bf n}')). 
\ee
Using $V_{\bf n}$-s we can define Lindblad-type coupling between the quantum 
system and the classical one.  To give our model more flavor,  we will
introduce also a quantum Hamiltonian that depends on the actual state
of the classical system; thus we define
\be
\left ( H\Psi\right) ({\bf n})=
H ({\bf n})\Psi ({\bf n}) = {\omega\over2}\sigma  ({\bf n})\Psi  ({\bf n}). 
\ee
Our coupling is now given by
\be
{\cal L}_{cq} \rho = -i[H, \rho ] +\int_{{\cal S}_{cl}} \left ( V_{\bf n}\rho
V_{\bf n}^{\star}-{1\over2}
\left\{ {V_{\bf n}}^{\star} V_{\bf n}, \rho\right\}\right) d\mu  ({\bf n}). 
\ee 
Notice that ${V_{\bf n}}^{\star}=
V_{\bf n}$ and $V_{\bf n}^2=\kappa\,  e ({\bf n})$. 
Now, 
$\int e ({\bf n}) d\mu  ({\bf n})$ being $U (2)$--invariant,  it must be
proportional to the identity.  Taking its trace we find that
$$\int e ({\bf n}) d\mu  ({\bf n}) = I/2 , $$  
and therefore
\be
{\cal L}_{cq}\rho=-i[H, \rho]+\int  V_{\bf
n}\rho V_{\bf n}\,  d\mu  ({\bf n}) -{\kappa\over 2} \rho . 
\ee 
Explicitly,  using the definition of $V_{\bf n}$,  we have
\be
 \left ({\cal L}_{cq}\rho\right) ({\bf n}) =
 -i{\omega\over2}[\sigma ({\bf n}), \rho ({\bf n})]+
 \kappa\int e ({\bf n'})\rho\left ({\bf
n'}  ({\bf n})\right)e ({\bf n'}) d\mu  ({\bf n'}) -{\kappa\over 2} \rho ({\bf n}). 
\ee
Notice that for each operator $a\in{\cal L} ({\cal H}_q)$
we have the following formula: \footnote{The formula 
is easily established for $a$
of the form $e ({\bf n'})$,  and 
then extended to arbitrary operators by linearity. }
\be
\int e ({\bf n})\,  a\,  e ({\bf n}) d\mu  ({\bf n})={1\over 6} (a+Tr (a)I). 
\ee

If $\omega=0$,  that is if we neglect the Hamiltonian part,  then
using this formula we can integrate
over $ {\bf n'}$ to get the effective Liouville operator for the
quantum state ${\hat\rho}=\int \rho ({\bf n}) d\mu  ({\bf
n})$: 
\be
 {\cal L}_{cq}{\hat\rho}={\kappa\over6}\left (I-2{\hat\rho}\right), 
\ee
with the solution
\be
{\hat\rho} (t)=\exp\left ({-\kappa t}/3\right)\rho (0)+{1-\exp\left ({-\kappa
t}/3\right)\over 2}I. 
\ee
It follows that,  as the result of the coupling,  the effective quantum state
undergoes a rather uninteresting time--evolution:  it
dissipates exponentially towards the totally mixed state ${I\over2}$,  and this
 does not depend on the initial state of the classical system. \\

Returning back to the case of non--zero $\omega$ we discuss now
the piecewise deterministic random process
of the two pure states ${\bf n}_q$ and
${\bf n}_{cl}$.  To compute it we proceed as in the previous example, 
with the
only change that now pure states of the quantum {\sl and} of the classical
system are parameterized by the same set - $S^2$ in our case.  To keep track of
the origin of each parameter we will use subscripts as in ${\bf n}_{cl}$ and
${\bf n}_{q}$.  As in the previous example each observable ${\bf A}$ of the
total
system determines a function $f_{\bf A}: S^2\times S^2\rightarrow\complex$ by
$$f_{\bf A}\left ({\bf n}_q, {\bf n}_{cl}\right)=Tr\left (e\left ({\bf
n}_q\right)A\left ({\bf n}_{cl}\right)\right). $$ 
The Liouville operator  ${\cal L}_{cq}$,  acting on observables,   can be then  
rewritten it terms of the functions  $f_{\bf A}$: 
 \be
 \begin{array}{ll}
\left ({\cal L}_{cq}f_{\bf A}\right)\left ({\bf n}_q, {\bf n}_{cl}\right)
& 
= \   (X_H\,  f) ({\bf n}_q, {\bf n}_{cl})  +
\cr\cr
& + \  \kappa\int p ({\bf n}_q, {\bf n'}_{q}) f_{\bf A}\left ({\bf n'}_{q}, 
{{\bf n'}_{q}} ({\bf n}_{q})\right)d\mu \left ({\bf n'}_{q}\right) +
\cr\cr
& - \ {\kappa\over 2} f_{\bf A}\left ({\bf n}_q, {\bf n}_{cl}\right), 
\end{array}
\ee
where $X_H$ is the Hamiltonian vector field
\be
 (X_H f) ({\bf n}_q, {\bf n}_{cl})={d\over dt}f\left (e^{-iH ({\bf n}_{cl})t}
\cdot {\bf n}_{q}
\right)\bigg\vert_{t=0} \  , 
\ee
and

\be
 p ({\bf n}, {\bf n'}) = Tr\left (e\left ({\bf n}\right)e\left ({\bf
n'}\right)\right)= (1+{\bf n\cdot n'})/2
\ee
is known as the transition probability between the two quantum states. \\
The PD process on ${\cal S}_q\times {\cal S}_{cl}$ can now be
described as follows.  Let ${\bf n}_q (0)$ and ${\bf n}_{cl} (0)$ be the
initial states of the quantum and of the classical system.  Then the
quantum system evolves unitarily according to the quantum Hamiltonian
$H ({\bf n}_{cl})$ until at a time instant $t_1$ a jump occurs.  The time
rate of jumps is governed by the homogeneous Poisson process with rate
$\kappa/2$.  The quantum state ${\bf n}_q (t_1)$ jumps to a new state
${{\bf n}'}_q$ with probability distribution $p ({\bf n}_q (t_1), {{\bf
n}'}_q)$ while ${\bf n}_{cl}$ jumps to ${{\bf n}'}_q  ({\bf n}_q (t_1))$
and the process starts again (see Fig. 2). 
\\ \\
\begin{figure}[hbt]
\begin{picture}(100,200)
\put (-125, -50)
{\includegraphics{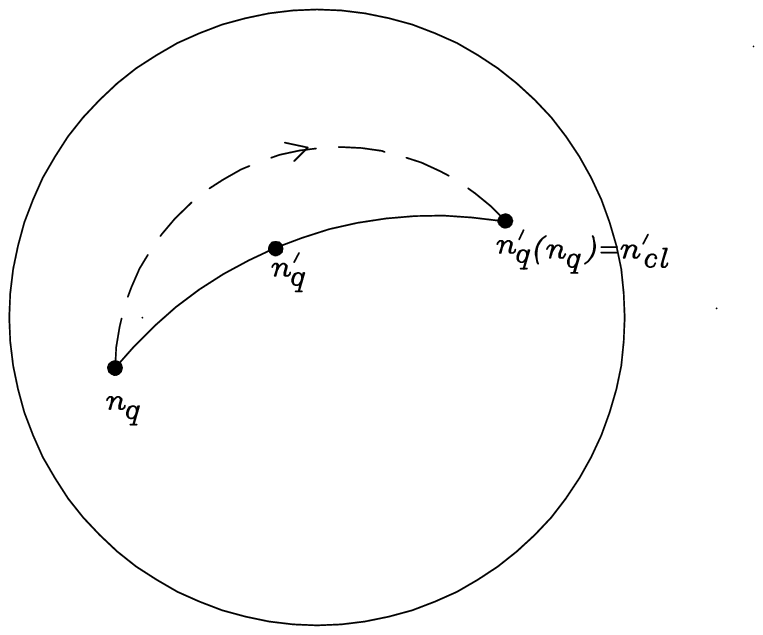}}
\end{picture}
\caption{The quantum state ${\bf n}_{q} (t_{1})$ jumps to a new state
${\bf n}^{\prime}_{q}$  with probability distribution 
$p ({\bf n}_{q} (t_{1}), {\bf n}^{\prime}_{q}$ while ${\bf n}_{cl}$
jumps to ${\bf n}^{\prime} ({\bf n}_{q} (t_{1}))$. }
\end{figure}
\vspace{1cm}
\newpage
\noindent
{\bf Acknowledgements: }
This paper is partially based on a series of publications that were 
done in a
collaboration with Ph.  Blanchard and M.  Modugno.  Thanks are due to
the Humboldt Foundation and Italian CNR that made this collaboration
possible. The financial support for the present work 
was provided by the Polish KBN grant No PB 1236  on one hand, and
European Community program "PECO" handled by the Centre de Physique
Theorique (Marseille Luminy) on the other.  
Parts of this work have been written while
I was visiting the CPT CNRS Marseille.  I would like to thank Pierre
Chiapetta and all the members of the Lab  
for their kind hospitality.  
 Thanks are due to the Organizers of the School
for  invitation and for providing travel support.  I would like to 
express
my  especially warm thanks to Robert Coquereaux for his constant
interest,  criticism and many fruitful discussions, 
 and for critical reading of much of this text. Thanks are also due
to Philippe Blanchard for reading the paper, many discussions and
for continuous, vital and encouraging interaction. 
 Last,  but not least,  thanks to Anna for her patience, 
 encouragement and for help with
the typing. 
\\ \\

\end{document}